\documentclass[amsmath,superscriptaddress,showpacs,prb,twocolumn,longbibliography]{revtex4-1}
\usepackage{bbm}
\usepackage{bm}
\usepackage{enumerate}
\usepackage{graphicx}
\usepackage[dvips]{epsfig}
\usepackage{epsf}
\usepackage[latin1]{inputenc}
\usepackage{mathtools}
\usepackage{amsmath}
\usepackage{amsfonts}
\usepackage{amssymb}
\usepackage{xcolor}
\usepackage{esint}

\bibpunct{[}{]}{;}{n}{}{}

\begin{document}
	\title{Superfluid Spin Transistor}
\author{Edward Schwartz}
\affiliation{Department of Physics and Astronomy and Nebraska Center for Materials and Nanoscience, University of Nebraska, Lincoln, Nebraska 68588, USA}

\author{Bo Li}
\affiliation{Department of Physics and Astronomy and Nebraska Center for Materials and Nanoscience, University of Nebraska, Lincoln, Nebraska 68588, USA}

\author{Alexey A. Kovalev}
\affiliation{Department of Physics and Astronomy and Nebraska Center for Materials and Nanoscience, University of Nebraska, Lincoln, Nebraska 68588, USA}
	
\begin{abstract}
We propose to use the Hall response of topological defects, such as merons and antimerons, to spin currents in two-dimensional magnetic insulator with in-plane anisotropy for identification of the Berezinskii-Kosterlitz-Thouless (BKT) transition in a transistorlike geometry. Our numerical results relying on a combination of Monte Carlo and spin dynamics simulations show transition from spin superfluidity to conventional spin transport, accompanied by the universal jump of the spin stiffness and exponential growth of the transverse vorticity current. We propose a superfluid spin transistor in which the spin and vorticity currents are modulated by changes in density of free topological defects, e.g., by injection of vorticity or by tuning the in-plane magnet across the BKT transition by changing the exchange interaction, magnetic anisotropy, or temperature. 
\end{abstract}

\maketitle
\section{Introduction}
Spintronics has emerged as a field addressing the need for low-power nanoelectronic devices by encompassing many recent developments in condensed matter physics~\cite{Hirohata2020,Fert2019}.
Spin degree of freedom, which is quantum in nature, also offers unique opportunities in studies of quantum materials~\cite{Han2020}. Berezinskii-Kosterlitz-Thouless (BKT) transition has been theoretically predicted in 2D XY model~\cite{berezinskii1971,Kosterlitz1973}, but substantial experimental progress has been achieved so far in nonmagentic systems such as thin films of superfluids/superconductors~\cite{PhysRevLett.40.1454,PhysRevLett.40.1727,PhysRevLett.47.534,PhysRevLett.44.291,PhysRevLett.107.217003} and superconducting arrays~\cite{PhysRevLett.47.1542,RevModPhys.56.431}.
Designing unambiguous experiments in magnetic systems is of high importance and spintronics methods seem to be suitable for identifying signatures of BKT transition~\cite{SciPostPhys,PhysRevLett.125.237204,PhysRevB.104.L020408}.  Recent experiments on magnetic van der Waals (vdW) materials such as NiPS$_3$, CrBr$_3$, and CrCl$_3$ further motivate theoretical research of magnetic BKT transition~\cite{Kim2019,Kim2019-1,BedoyaPinto2021}.

The difficulty in detecting magnetic BKT transition arises due to absence of the long-range order parameter characteristic to phase transitions within the Landau paradigm~\cite{PhysRevLett.17.1133,PhysRev.158.383}. Instead, the low temperature region below BKT transition is associated with appearance of bound topological defects. Above BKT transition, the behavior is determined by exponentially increasing density of unbound topological defects. Recent theoretical works have addressed behavior of spin and charge currents in the vicinity of magnetic BKT transition~\cite{SciPostPhys,PhysRevLett.125.237204,PhysRevB.104.L020408}. The behavior of spin current is of particular interest due to apparent analogy to superfluid transport described by the $U(1)$ phase gradient~\cite{SciPostPhys}. Spin superfluid transport has been proposed in collinear~\cite{PhysRevLett.112.227201,PhysRevB.90.094408,PhysRevLett.115.237201,PhysRevLett.87.187202,PhysRevB.95.144432,PhysRevLett.116.117201,PhysRevB.96.134434,PhysRevB.103.144412} and noncollinear~\cite{PhysRevB.103.L060406,PhysRevB.103.104425} magnets and realized in recent experiments~\cite{Stepanov2018,Yuan2018}. 

Transport signatures can be used for identification of BKT transition, e.g., as has been established in literature on superconducting systems.
While applying these ideas to magnetic insulating systems one needs to be aware of differences such as the nonconservation of spin currents~\cite{SciPostPhys}, the coupling to phonons, and the presence of merons rather than vortices as topological defects. This warrants both identification of new physics and development of new methods for studying relevant magnetic systems.  

\begin{figure}
\centering
\includegraphics[width=\linewidth]{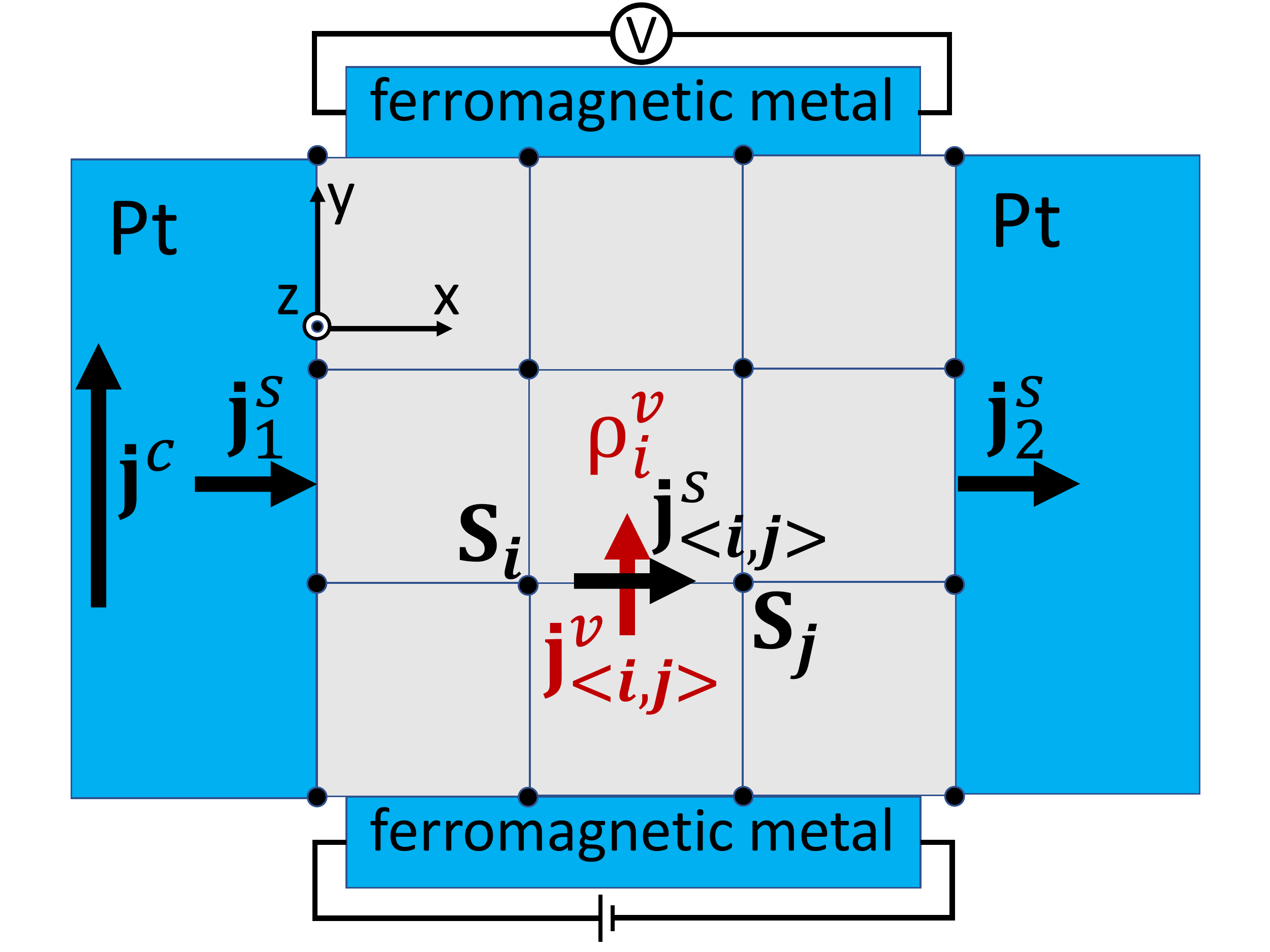}
\caption{(Color online) A schematic plot of the system. Spin current is injected into an easy-plane magnet using a heavy metal such as Pt. Spin and vorticity currents, and vorticity density are shown for a single plaquette. The vorticity current can be detected by the top ferromagnetic metal contact magnetized along the z-axis. The spin current can be modulated by tuning in-plane magnet below or above BKT transition via modifications of exchange interactions and anisotropy, e.g., by the gate modulated electrical field-induced strain or direct electrical-field-induced modifications of exchange interactions and anisotropy (the gate is assumed in the $x-y$ plane). Alternatively, the spin superfluid current can be modulated by injection of vorticity by the bottom ferromagnetic metal contact magnetized along the z-axis via charge current flow through the ferromagnet.}
\label{fig:spinvorticity}
\end{figure}

In this work, we propose to use transport signatures of the Hall response of topological defects, such as merons and antimerons, as a hallmark of BKT transition. Starting from the Landau-Lifshitz-Gilbert type phenomenology, we derive the constitutive phenomenological relations~\cite{SciPostPhys} describing the vorticity and spin currents in insulating magnetic systems in the presence of Gilbert damping and spin nonconservation.
We confirm these relations and a possibility of long range spin superfluidity below BKT transition by employing a combination of Monte Carlo and spin dynamics simulations. We propose a superfluid spin transistor (see Fig.~\ref{fig:spinvorticity}) in which spin superfluid current is modulated by tuning in-plane magnet below or above BKT transition, e.g., by changing the exchange interaction, magnetic anisotropy, or temperature. Spin current injected through lead can propagate in the spin superfluidity regime below BKT transition while spin current will exponentially decay above BKT transition. The presence of transverse vorticity current can be detected using the top ferromagnetic contact in Fig.~\ref{fig:spinvorticity}. Alternatively, the spin superfluid current below BKT transition can be modulated in Fig.~\ref{fig:spinvorticity} by injection of vorticity by the bottom ferromagnetic metal magnetized along the z-axis via charge current flow through the ferromagnet.

\section{Spin and vorticity transport from spin dynamics}	
We consider a magnetic insulator with the in-plane magnetic anisotropy at finite temperature described by the Hamiltonian:
\begin{equation} \label{eq:ham}
    {\cal H} =-J \sum_{\left< i,j \right>} (S^x_i S^x_j+S^y_i S^y_j+\lambda S^z_i S^z_j)+2 J \beta \sum_i (S^z_i)^2 ,
\end{equation}
where $J>0$~\cite{Note} is the ferromagnetic exchange interaction, and dimensionless parameters $\lambda$ ($0\leq\lambda<1$) and $\beta$ describe the magnetic anisotropy. Such anisotropic interactions can be realized in 2D vdW magnets~\cite{PhysRevLett.124.017201}. Note that Eq.~\eqref{eq:ham} can lead to both in-plane vortex-like solutions as well as stable meron-like solutions~\cite{PhysRevB.39.11840}. The spin dynamics is determined by the discrete version of the Landau-Lifshitz-Gilbert equation:
\begin{equation}\label{eq:LLG}
     s(1-\alpha {\bf S}_i\times)\,\partial_t {\bf S}_i={\bf S}_i\times {\bf H}_i,  
\end{equation}
where $s$ is the spin density, ${\bf H}_i={\bf H}_i^\text{eff}+{\bf H}_i^\text{th}$ includes the effective field ${\bf H}_i^\text{eff}=\frac{J}{a^2}[\sum_{j\in N(i)} (S^x_j,S^y_j,\lambda S^z_j)- (0,0,4\beta S^z_i)]$ with $N(i)$ denoting the nearest neighbours, ${\bf H}_i^\text{th}$ is the thermal field describing the Langevin force, and $\alpha$ is the Gilbert damping. Note that the Gilbert damping term and the Langevin force could correspond to weak coupling to external degrees of freedom such as phonons. For simplicity we assume a square lattice; however, our approach can be generalized to other lattices.   

In the absence of coupling to external degrees of freedom, we can define spin current for Hamiltonian \eqref{eq:ham} according to continuity equation
$\partial_t \rho^s +{\boldsymbol \nabla} \cdot {\bf j}^s=0$. Using the finite difference approximation with $\rho^s_i=s S_i^z$, one can show that the spin current associated with each bond can be expressed as ${\bf j}^s_{\left< i,j \right>}=\frac{J}{a^2}({\bf r}_i-{\bf r}_j) ({\bf z}\cdot{\bf S}_i\times {\bf S}_j)$ (see Fig.~\ref{fig:spinvorticity}). Using Eq.~\eqref{eq:LLG}, we can write down modification of continuity equation due to non-conservation of spin in discrete form:
\begin{equation}\label{eq:spin}
    \partial_t \rho^s_i +{\boldsymbol \nabla}_i \cdot {\bf j}^s_{\left< i,j \right>}=\alpha s\,{\bf z}\cdot{\bf S}_i\times\partial_t{\bf S}_i,
\end{equation}
where ${\boldsymbol \nabla}_i \cdot {\bf f}_{\left< i,j \right>}=\frac{1}{a^2}\sum_{j\in N(i)}({\bf r}_i-{\bf r}_j){\bf f}_{\left< i,j \right>}$. In the long wavelength limit, Eq.~\eqref{eq:spin} can be written as $\partial_t \rho^s+\boldsymbol \nabla {\bf j}^s=-\rho^s/\tau$ as follows from a relation, ${\bf z}\cdot{\bf S}\times\partial_t{\bf S}\propto \rho^s$, applicable to dynamics in the vicinity of the in-plane configuration. 

For each plaquette labeled by left and lower lattice site, we can define the vorticity density according to $\rho^v_i=\sum_{\left< i,j \right>\in {\cal P}(i)} \frac{1}{2 \pi a^2}({\bf z}\cdot{\bf S}_i\times {\bf S}_j) $ where ${\cal P}(i)$ denotes all edges of plaquette $i$ and the ordering of edge indices corresponds to ${\bf r}_i-{\bf r}_j$ pointing along the counterclockwise walk around the plaquette (see Fig.~\ref{fig:spinvorticity}). Using these definitions, we can immediately obtain a discretized relation between the vorticity and the spin current:
\begin{equation}\label{eq:vdensity}
   \sum_{i\in S} \rho^v_i a^2=\frac{1}{2 \pi J}\ointctrclockwise_{\partial S} {\bf j}^s_{\left< i,j \right>} d{\bf l},
\end{equation}
where $\partial S$ is defined by a set of bonds forming a closed path.
Note that Eq.~\eqref{eq:vdensity} in the long wavelength limit becomes $\rho^v={\bf z}\cdot (\boldsymbol \nabla \times {\bf j}^s)/2\pi J$, which is analogous to relation between the Cooper charge current and the superconducting vorticity in the theory of superconducting films.

In addition to spin current, we also associate a conserved vorticity current with each bond. The vorticity current is defined for each bond as ${\bf j}^v_{\left< i,j \right>}=\frac{1}{2 \pi a^2}{\bf z}\times({\bf r}_i-{\bf r}_j) [{\bf z}\cdot({\bf S}_i-{\bf S}_j)\times \partial_t ({\bf S}_i+{\bf S}_j)]$~\cite{PhysRevB.102.224433}. By using the discrete representation, one can confirm the following relation
\begin{equation}\label{eq:vcurrent}
    {\bf j}^v_{\left< i,j \right>}=\frac{1}{2 \pi J} {\bf z}\times\left(\partial_t {\bf j}^s_{\left< i,j \right>}+J\boldsymbol\nabla_{\left< i,j \right>}[{\bf z}\cdot{\bf S}\times\partial_t{\bf S}]\right),
\end{equation}
where $\boldsymbol \nabla_{\left< i,j \right>}f=(f_i-f_j)({\bf r}_i-{\bf r}_j)/a^2$. Note that the same combinations related to spin density appear in the right-hand sides of Eqs.~\eqref{eq:spin} and \eqref{eq:vcurrent}. Thus, Eq.~\eqref{eq:vcurrent} in the long wavelength limit becomes ${\bf j}^v=\frac{1}{2 \pi J} {\bf z}\times ( \partial_t {\bf j}^s+v^2\boldsymbol\nabla \rho^s)$, which is analogous to relation between the vorticity current, the Cooper charge current, and the charge density in the description of superconducting films. The continuous versions of Eqs.~\eqref{eq:spin}, \eqref{eq:vdensity}, and \eqref{eq:vcurrent} have been written in Ref.~\cite{SciPostPhys} by employing analogies between superconducting films and easy-plane magnets.

We assume that spin is injected into 2D magnetic insulator by using the spin Hall effect at the left edge and detected by using the inverse spin Hall effect at the right edge, see Fig.~\ref{fig:spinvorticity}. Alternatively, we also consider a setup in which spin currents with opposite polarizations are injected from the left and right edges while a top ferromagnetic contact (Fig.~\ref{fig:spinvorticity}) is used to measure the vorticity current. Note that we only excite dynamics close to the in-plane configuration by choosing sufficiently weak injected currents. 

To model the injection of spin current, we assume that spins on a square lattice closest to the injector will experience the spin-orbit torque, 
\begin{equation}\label{eq:SO}
\boldsymbol \tau_{so}=\frac{\vartheta j^c}{a} {\bf m}\times[{\bf m}\times{\bf z}],
\end{equation}
as well as an additional Gilbert damping, $\alpha^\prime=\hbar g^{\uparrow\downarrow}/4\pi$, where $g^{\uparrow\downarrow}$ is the effective spin mixing conductance and $\vartheta$ is the effective spin Hall coefficient, i.e., $j^s=\vartheta j^c$. When a contact is only used as a spin current detector, the closest spins will also experience an additional Gilbert damping, $\alpha^\prime$. 

\section{Phenomenological description}
To understand the behavior of 2D magnetic insulator across BKT transition, we study the spin and vorticity current responses.
To describe the system phenomenologically in the long wavelength limit, one needs to supplement Eqs.~\eqref{eq:spin} and \eqref{eq:vcurrent} with the phenomenological equation describing the Magnus force on topological defects~\cite{PhysRevLett.124.157203,SciPostPhys}:
\begin{equation}\label{eq:magnus}
    {\bf j}^v=\mu n_f {\bf j}^s\times {\bf z},
\end{equation}
where $n_f$ is the total density of combined free topological defects with positive and negative vorticity, and $\mu$ is their mobility.
The spin and vorticity current responses undergo changes as temperature changes across BKT transition. At $T>T_\text{BKT}$, the behavior is dominated by the temperature dependence of the free topological defect density, i.e., $ n_f^> \propto \exp(-2b/\sqrt{T/T_{BKT}-1})$~\cite{PhysRevLett.40.783}. At $T<T_\text{BKT}$, there are no free topological defects; however, the presence of spin current can break some bound pairs of opposite vorticity resulting in the density~\cite{SciPostPhys} $n_f\propto \exp(-\Delta F/k_B T)$ where $\Delta F \approx \pi \Tilde{K} \ln({\cal J}^s/j^s)$ is the free energy barrier for the process of unbinding a pair and ${\cal J}^s$ is a phenomenological parameter describing modification of the free energy due to the Magnus force, which results in $n_f^<\propto (j^s/{\cal J}^s)^{\pi \Tilde{K}/k_B T}$~\cite{Halperin2010,SciPostPhys}. Here $\Tilde{K}$ is the
vortex-renormalized spin stiffness for which at BKT temperature we have a relation, $\pi \Tilde{K}/2=k_B T_\text{BKT}$.

Combining Eqs.~\eqref{eq:spin}, \eqref{eq:vcurrent}, and \eqref{eq:magnus} one can obtain equations for spin current in DC limit for $T>T_\text{BKT}$:
\begin{equation}\label{eq:diffusion}
    {\bf j}^s=\left(\lambda^{^>}\right)^2 \boldsymbol \nabla \left(\boldsymbol \nabla\cdot {\bf j}^s\right),
\end{equation}
where $\lambda^{^>}=(2\pi \alpha s \mu n_f^>)^{-1/2}$, and for $T<T_\text{BKT}$~\cite{SciPostPhys}: 
\begin{equation}\label{eq:SS}
    \left(\frac{j^s}{{\cal J}^s}\right)^{\pi\Tilde{K} /k_B T}\frac{{\bf j}^s}{{\cal J}^s}=\left(\lambda^{^<}\right)^2 \boldsymbol \nabla \left(\boldsymbol \nabla\cdot \frac{{\bf j}^s}{{\cal J}^s}\right),
\end{equation}
where $\lambda^{^<}=(j^s/{\cal J}^s)^{\pi\Tilde{K} /2 k_B T}(2\pi \alpha s \mu n_f^<)^{-1/2}$. These equations can be supplemented with the boundary conditions for our setup in Fig.~\ref{fig:spinvorticity} by using Eqs.~\eqref{eq:LLG}, \eqref{eq:spin}, and \eqref{eq:SO}:
\begin{align}\label{eq:boundary}
    j^s(0)&=j^s_1+\Tilde{g}\partial_x j^s |_{x=0},\\
     j^s(L)&=j^s_2-\Tilde{g}\partial_x j^s |_{x=L}, \label{eq:boundary1}
\end{align}
where $\Tilde{g}=a \alpha^\prime/\alpha$ and we also consider a possibility of spin current injection from the right lead in Fig.~\ref{fig:spinvorticity}.  

The algebraic decay of spin current at large $L$ for $T<T_\text{BKT}$ can be obtained by analyzing Eq.~\eqref{eq:SS} (see Appendix~A).
When $\Tilde{g}=0$ and $j_1^s=0$ by integrating Eq.~\eqref{eq:SS}, we obtain expression for the reduced spin current $j = j^s/{\cal J}^s $ expressed in terms of the inverse function $j(x)=f^{-1}(x)$:
\begin{align}\label{eq:spin-current}
    f(j)=\frac{j}{\partial_xj|_{x=0}} \prescript{}{2}{F}_1\Big( \frac{1}{2}, \frac{1}{2\nu}, 1 + \frac{1}{2\nu}; - \frac{j^{2\nu}}{k \nu}\Big) ,
\end{align}
where $\nu=1+\pi\Tilde{K} /2 k_B T$, $k = (\lambda^<)^2(\partial_xj|_{x=0})^2$, and $\prescript{}{2}{F}_1(\dots)$ stands for the hypergeometric function. The boundary condition $j(L)=j_2^s/{\cal J}^s$ can be used to find $\partial_xj|_{x=0}$. For small $\Tilde{g}/L$, we treat $\Tilde{g}/L$ as a small parameter in the first order expansion $j(x)=j^{(0)}(x)+(\Tilde{g}/L)j^{(1)}(x)$ to find the spin current injected in the opposite lead, $j(0)=\Tilde{g}\partial_x j^{(0)}|_{x=0}$. When $j_2^s=0$ we replace $x$ by $L-x$ in Eq.~\eqref{eq:spin-current}, use boundary condition $j(0)=j_1^s/{\cal J}^s$, and obtain $j(L)=-\Tilde{g}\partial_x j^{(0)}|_{x=L}$. Note that Eq.~\eqref{eq:spin-current} shows algebraic asymptotic behavior for the spin current detected in the lead opposite to the injection lead, i.e., $j^s[L]\propto\Tilde{g} L^{\nu/(1-\nu)}$ at large $L$~\cite{SciPostPhys}. More general boundary conditions~\eqref{eq:boundary} are discussed in Appendix~A. The exponential decay of spin current, $j^s[L]\propto\exp(-L/\lambda^{^>})$, for $T>T_\text{BKT}$ can be easily obtained by solving  Eq.~\eqref{eq:diffusion} describing spin diffusion.

We characterize the vorticity response using the averaged vorticity Hall current defined by relation:
\begin{equation}\label{eq:VHE}
    J^v=\frac{1}{L}\int_0^L j^v dx .
\end{equation}
This vorticity response can be detected by a ferromagnetic metallic contact in Fig.~\ref{fig:spinvorticity}. In particular, an exchange coupling between ferromagnetic metal and magnetic insulator will result in electromotive force~\cite{PhysRevB.99.180402}, $\boldsymbol\varepsilon =\eta {\bf j}^v \times {\bf M}$, here $\eta$ is a phenomenological parameter quantifying the strength of coupling and $\bf M$ is the magnetization along the z-axis.
This translates into a measurable voltage in the ferromagnetic contact given by $V=\eta L M J^v$.

For simplicity, we assume $\Tilde{g}=0$ as it leads to a small correction for small $\Tilde{g}$. We assume injection of spin current from both sides in Fig.~\ref{fig:spinvorticity}, i.e., $j_1^s=j_2^s$.
By integrating Eqs.~\eqref{eq:diffusion} and \eqref{eq:SS} and using Eq.~\eqref{eq:spin}, we obtain the vorticity Hall responses for $T<T_\text{BKT}$:
\begin{align}
    J^v=\frac{{\cal J}^s}{\pi\alpha s }\frac{\sqrt{[j(0)]^{2\nu}-[j(L/2)]^{2\nu}}}{\sqrt{\nu} \lambda^{^<} L}
    &\stackrel{\frac{L j}{\lambda^{^<}}\gg 1}{\approx}\frac{{\cal J}^s}{\pi\alpha s} \frac{[j(0)]^{\nu}}{\sqrt{\nu}\lambda^{^<} L}\\
    &\stackrel{\frac{L j}{\lambda^{^<}}\ll 1}{\approx}\frac{{\cal J}^s}{\pi\alpha s} \frac{[j(0)]^{2\nu-1}}{2(\lambda^{^<})^2}\,,
\end{align}
and for $T>T_\text{BKT}$:
\begin{align}\label{eq:vort-tot}
    J^v=\frac{j^s(0)}{\pi\alpha s }\frac{1}{\lambda^{^>} L\coth(L/2\lambda^{^>})}
    &\stackrel{\frac{L }{\lambda^{^>}}\gg 1}{\approx}\frac{1}{\pi\alpha s} \frac{j^s(0)}{\lambda^{^>} L}\\
    &\stackrel{\frac{L }{\lambda^{^>}}\ll 1}{\approx}\frac{1}{\pi\alpha s} \frac{j^s(0)}{2(\lambda^{^>})^2}\,.
\end{align}
These equations demonstrate that the vorticity Hall response dependence on injected spin current changes across BKT transition from linear to nonlinear behavior with the power factor determined by the spin stiffness. Furthermore, it should be possible to determine the transition temperature by fitting the vorticity Hall response to temperature dependence determined by $\lambda^{^>}$ above BKT transition.
\begin{figure}
\centering
\includegraphics[width=0.7\linewidth]{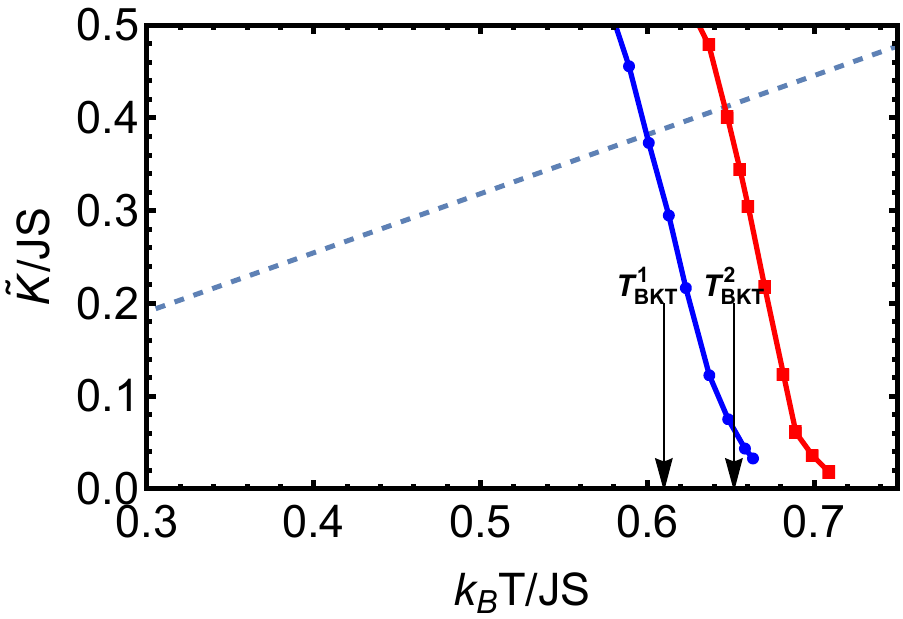}
\caption{(Color online) Rernormalized spin stiffness, $\Tilde{K}$, in the vicinity of BKT transition. The blue line corresponds to the anisotropy $\beta=0.05$. The red line corresponds to the anisotropy $\beta=0.1$. Arrows correspond to estimated BKT temperatures. The dashed line, $2 k_B T/\pi$, can be used to estimate the position of universal jump of the spin stiffness.}
\label{fig:stiffness}
\end{figure}

\begin{figure}
\centering
\includegraphics[width=0.49\linewidth]{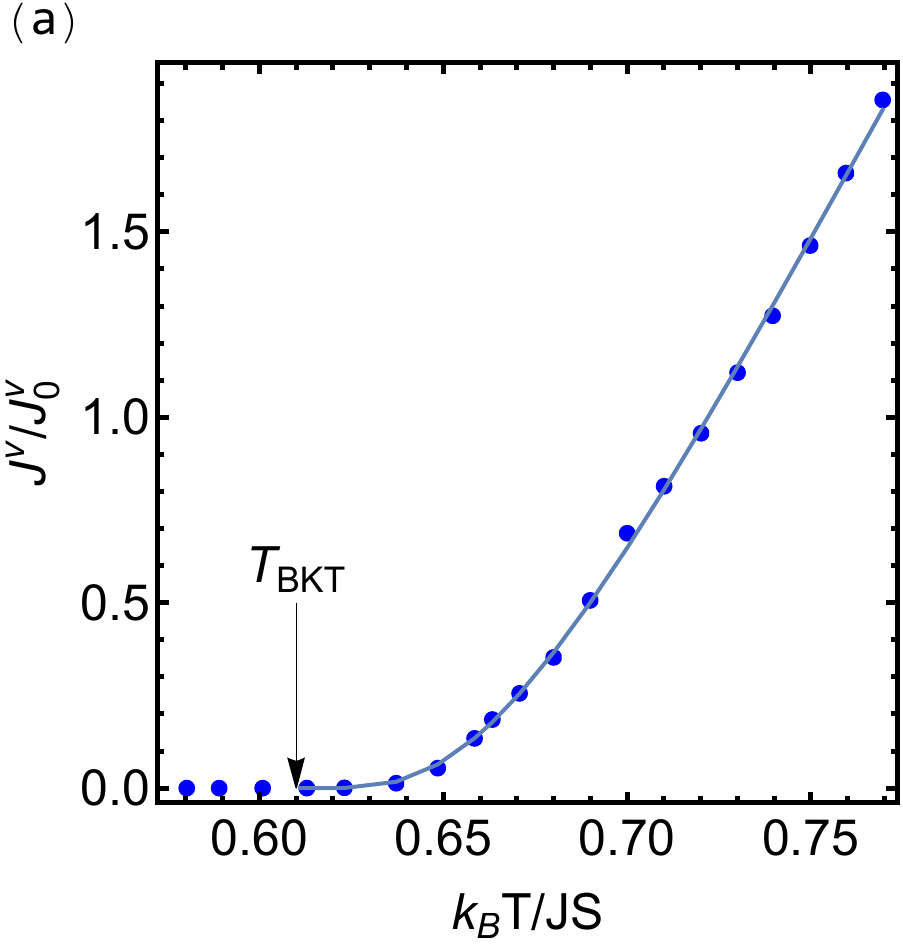}
\includegraphics[width=0.49\linewidth]{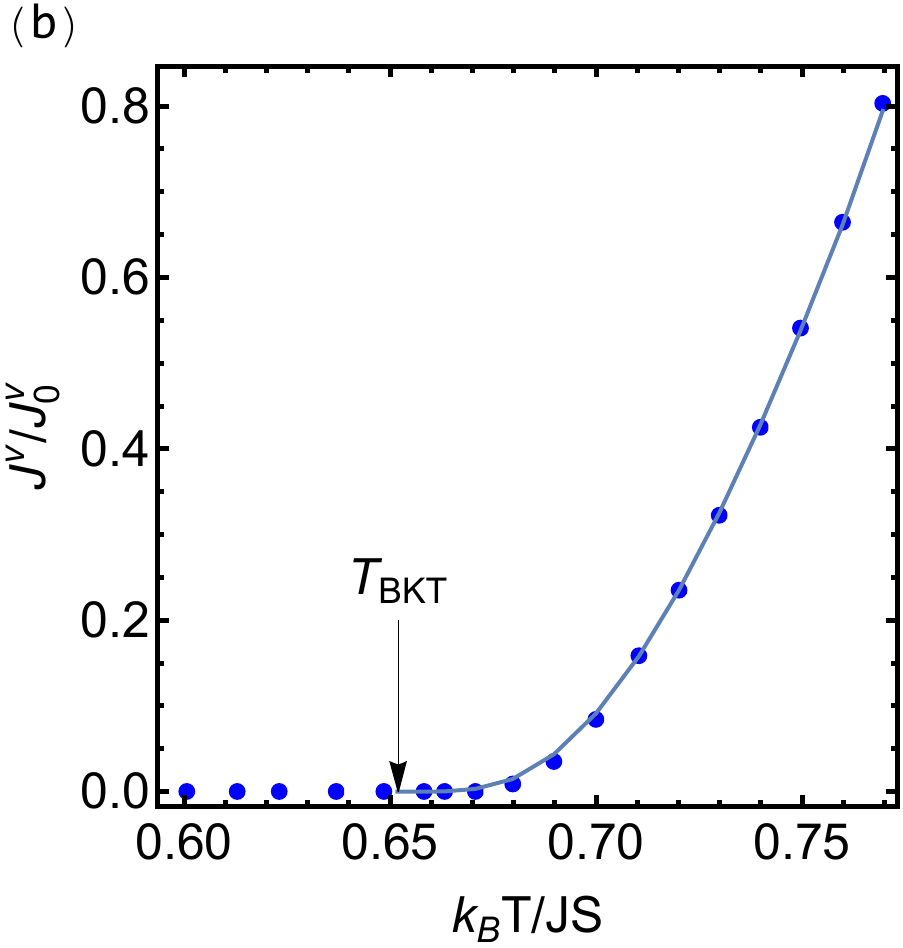}
\caption{(Color online) Circles show numerical results for the vorticity Hall response as a function of temperature where $J_0^v=J/(2\pi\alpha s a^2 L)$.  Lines show fit to Eq.~\eqref{eq:vort-tot}. (a) Fitting for anisotropy $\beta=0.05$ leads to estimate $k_B T_\text{BKT}/J\approx0.61\pm 0.01$. (b) Fitting for anisotropy $\beta=0.1$ leads to estimate $k_B T_\text{BKT}/J\approx0.65\pm0.01$.}
\label{fig:vort}
\end{figure}
\begin{figure}
\centering
\includegraphics[width=0.75\linewidth]{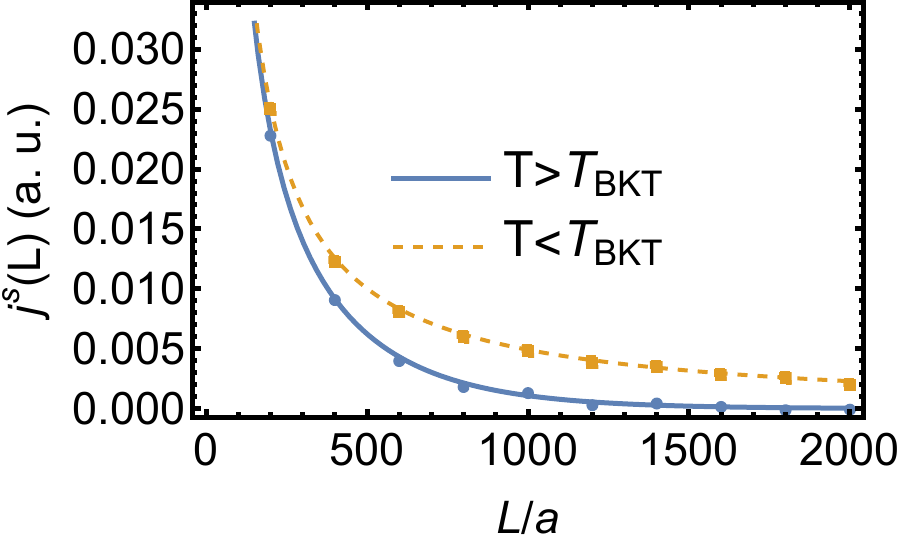}
\caption{(Color online) Spin current $j^s(L)$ for different system sizes when spin current $j^s(0)$ is injected on the left. We consider in-plane anisotropy $\beta=0.1$ and take $T=0.62 J/k_B<T_\text{BKT}$ for the upper plot and $T=0.71 J/k_B>T_\text{BKT}$ for the lower plot. Lines are the fit to Eq.~\eqref{eq:spin-current}. Curves demonstrate crossover from exponential to algebraic decay.}
\label{fig:spin}
\end{figure}

\section{Numerical results and discussions}
The spin dynamics simulations corresponding to Eq.~\eqref{eq:LLG} are performed from $100$ initial configurations using Mumax3~\cite{Vansteenkiste2014} code with the Gilbert damping $\alpha=0.001$. The initial configurations for spin dynamics simulations are obtained by using the feedback-optimized
parallel tempering Monte Carlo simulations~\cite{Katzgraber2006} on  $200\times200$ lattices with periodic boundary conditions where $10^6$ metropolis updates per spin have been performed to equilibrate the system (see Appendix~B). Averaging over initial configurations and time averaging are performed after we reach a steady state in the spin dynamics simulations (see Appendix~C for a snapshot of magnetic texture obtained by the spin dynamics simulation). As an output, we calculate averaged quantities $\left< j^s \right>_\text{av}$ and $ \left< \rho^s \right>_\text{av}$, where the latter can be related to the vorticity current using Eq.~\eqref{eq:vcurrent}. To obtain results in Figs.~\ref{fig:stiffness} and \ref{fig:vort}, we perform spin dynamics simulations on lattices with the longitudinal size of $1600$ sites, periodic boundaries, and boundary conditions in Eqs.~\eqref{eq:boundary} and \eqref{eq:boundary1} with $\Tilde{g}=0$, $j_1^s=j_2^s$. In Fig.~\ref{fig:spin}, we use a variable longitudinal size of up to $4000$ sites with spin injection only on the left side (see Fig.~\ref{fig:spinvorticity}).

To obtain the renormalized spin stiffness in the vicinity of BKT transition in Fig.~\ref{fig:stiffness}, we use Eq.~\eqref{eq:magnus} and perform numerical simulations for different values of spin and vorticity currents. In our calculations, we concentrate on effects of magneto-crystalline anisotropy $\beta$. The exchange anisotropy $\lambda$ (not shown) gives qualitatively similar results, also consistent with other approaches in the limit of large $S$~\cite{PhysRevB.45.2883,PhysRevB.52.10221}. Due to finite size effects in Fig.~\ref{fig:stiffness}, we observe a gradual decay of the spin stiffness instead of the universal jump~\cite{PhysRevB.49.12071}. 
In Fig.~\ref{fig:vort}, we study the behavior of the vorticity Hall response above BKT transition. By fitting to Eq.~\eqref{eq:vort-tot}, we estimate BKT temperatures for the in-plane anisotropies $\beta=0.05$ and $\beta=0.1$ where the latter estimate is in agreement with previous results~\cite{PhysRevB.99.180502}.
In Fig.~\ref{fig:spin}, we numerically study propagation of spin current in in-plane magnet. We inject spin on the left side and assume $\Tilde{g}=0$. Up to the leading order for small $\Tilde{g}$, we can estimate spin current on the opposite side as $j^s(L)=-\Tilde{g}\partial_x j^s|_{x=0}$. This spin current is plotted in Fig.~\ref{fig:spin} as a function of system size $L$, along with a line fit to result of Eq.~\eqref{eq:spin-current}. We observe a clear crossover from diffusive to spin superfluid transport.

\section{Conclusions}
We have developed a description of 2D magnetic insulators based on a combination of Monte Carlo and spin dynamics simulations in the presence of Gilbert damping and spin nonconservation.
Electrical and strain control of magnetism in vdW magnets, e.g., by using electrical-field-induced strain or direct electrical-field-induced modifications of exchange interactions and anisotropy, is an active topic of research~\cite{PhysRevB.98.144411,Jiang2018,Burch2018,Verzhbitskiy2020,aelm.201900778,Jiang2021}. Through renormalization of exchange and anisotropy, a change in compressive strain of $0.5\%$ is sufficient to change BKT temperature by $10$K in CrCl$_3$ according to recent DFT calculations~\cite{PhysRevLett.127.037204}. Such strain modulation is available, e.g., by using a piezoelectric strain cell~\cite{Cenker2022} potentially allowing electrical control of spin and vorticity flows in a transistor-like geometry by tuning an in-plane magnet across BKT transition.
Furthermore, the same setup can be used for measuring the Hall response of topological defects, such as merons and antimerons, in a form of vorticity current that exhibits changes across BKT transition. Alternatively, the spin superfluid current can be modulated by controlling the density of free topological defects via injection of vorticity by the ferromagnetic metal. Finally, the changes, associated with crossover from spin superfluidity to conventional spin transport, universal jump of the spin stiffness, and
exponential growth of the transverse vorticity current, can be used for identifying the presence of BKT transition in a transport experiment relying on techniques commonly used in spintronics.  

\begin{acknowledgments}
This work was supported by the U.S. Department of Energy, Office of Science, Basic Energy Sciences, under Award No. DE-SC0021019. Part of this
work was completed utilizing the Holland Computing
Center of the University of Nebraska, which receives support
from the Nebraska Research Initiative.
\end{acknowledgments}

\appendix

\section{General Solution for $j^s(x)$}

For $T<T_{BKT}$ the (one dimensional) transport of the spin current $j^s$ is described by equation (10):

\begin{equation}
    \Big(j\Big)^{1 + 2T_{BKT}/T} = (\lambda^<)^2\partial_x^2j
\end{equation}

Where $j = j^s/{\cal J}^s $.
Multiplying both sides by $\partial_x j$ and integrating between $x$ and and arbitrary point $x_0$ gives a separable, first order differential equation:

\begin{equation}
    \frac{1}{2\nu}(j^{2\nu}-j^{2\nu}|_{x=x_0}) = \frac{1}{2}(\lambda^<)^2[(\partial_xj)^2-(\partial_xj|_{x=x_0})^2]
\end{equation}

Which after separation results in:

\begin{equation}
    \int_{j(x_0)}^{j(x)}\frac{dj}{\sqrt{\frac{j^{2\nu}}{k^2}+1}} = \pm \frac{|k|}{\sqrt{\nu} \lambda^<}(x - x_0)
\end{equation}

Where $k^2 = \nu (\lambda^<)^2(\partial_xj|_{x=x_0})^2 - j^{2 \nu}|_{x=x_0}$. The integral on the left may be calculated exactly in terms of the hypergeometric function:

\begin{equation}
    \int \frac{dx}{\sqrt{\frac{x^\alpha}{c}+1}} = x\prescript{}{2}{F}_1\Big(1/2, \frac{1}{\alpha}, 1 + \frac{1}{\alpha}; -\frac{x^\alpha}{c}\Big) + const
\end{equation}

\begin{equation}
    \prescript{}{2}{F}_1(a,b,c;z) = \frac{\Gamma(c)}{\Gamma(a)\Gamma(b)}\sum_{n=0}^{\infty } \frac{\Gamma(n+a)\Gamma(n+b)}{\Gamma(n+c)n!}z^n
\end{equation}

Which leaves us with a transcendental equation:

\begin{equation}
    \Big[ j \prescript{}{2}{F}_1\Big( 1/2, \frac{1}{2\nu}, 1 + \frac{1}{2\nu}; - \frac{j^{2\nu}}{k^2}\Big) \Big] \Big|_{j=j(x_0)}^{j(x)} = \pm \frac{|k|}{\sqrt{\nu} \lambda^<}( x - x_0)
\end{equation}

Which may be solved numerically for $j(x)$. Note that there is still some ambiguity in the choice of sign on the right hand side. This may be resolved by taking the behavior of the function on the left into account. 
\begin{figure}
\centering
\includegraphics[width=0.8\linewidth]{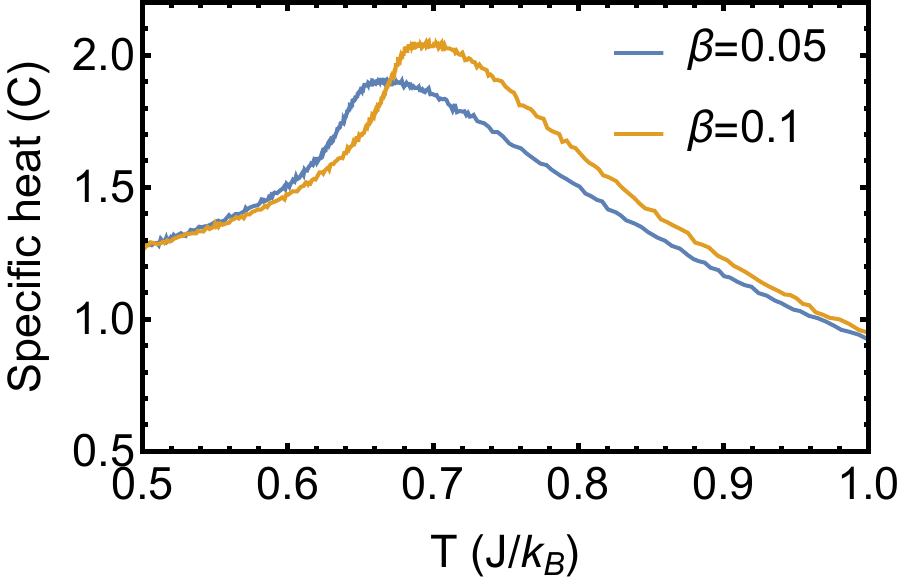}
\caption{(Color online) Specific heat $C$ in units of $k_B$ per spin as a function of temperature in units of $J/k_B$ for different magnetic anisotropies expressed in terms of dimensionless parameter $\beta$.}
\label{fig:cv}
\end{figure}

From the derivative, we can see that for finite $j>0$ the function $j \prescript{}{2}{F}_1( ... )$ is a monotonically increasing function of $j$. Therefore in regions where $j$ is an increasing (decreasing) function of $x$ the right side must be as well, and the correct solution requires the $+$ ($-$) sign.
If both regions with with the $+$ and $-$ sign are present, the general solution can be found by treating these regions separately and then linking the solutions.

\section{Specific heat calculations for different anisotropies}
In Fig.~\ref{fig:cv}, we calculate the specific heat using the feedback-optimized
parallel tempering Monte Carlo simulations~[39] on $200\times200$ lattices with periodic boundary conditions where $10^6$ metropolis updates per spin have been performed to equilibrate the system. As expected, the specific heat shows a peak that is shifted with respect to the position of BKT transition. We observe agreement with results reported in Refs.~[40] and [43].

\section{Magnetic textures}
\begin{figure}
\centering
\includegraphics[width=0.8\linewidth]{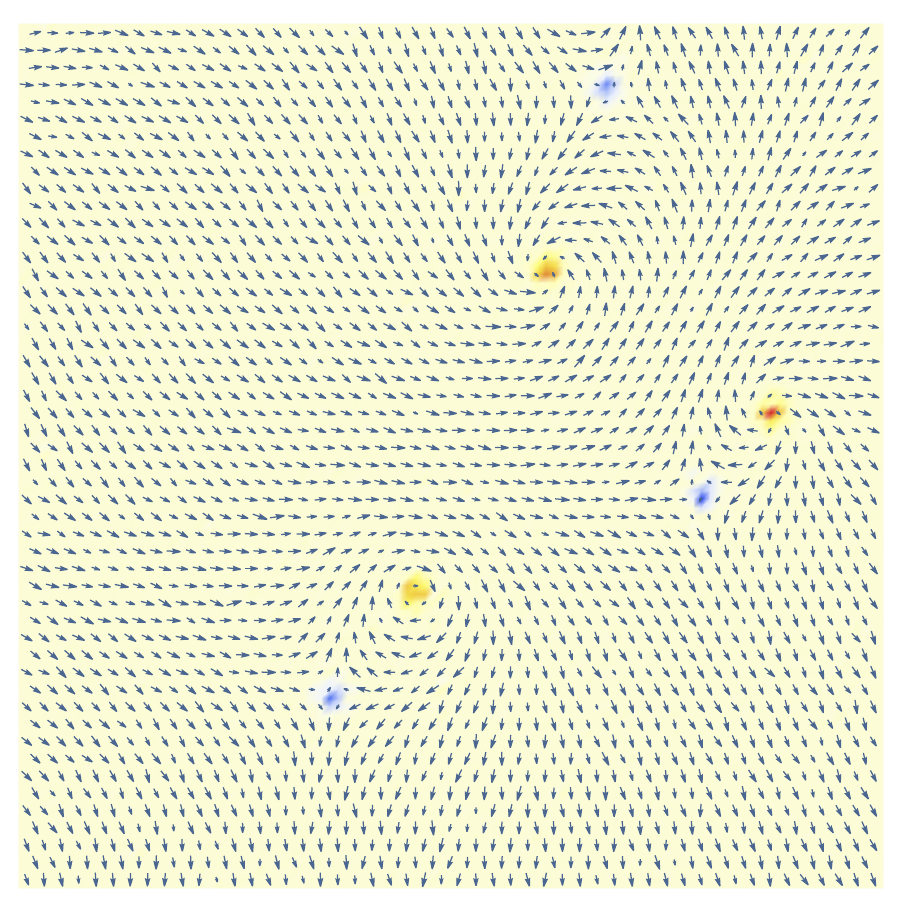}
\caption{(Color online) A snapshot containing three meron-antimeron pairs obtained by spin dynamics simulations at $T=0.2J/k_B$ for anisotropy $\beta=0.1$. Color indicates the sign and magnitude of vorticity density.}
\label{fig:merons}
\end{figure}

To visually assess the presence of merons and antimerons, we can take snapshots from spin dynamics simulations. Locations of defects can be further determined by calculating the vorticity density:
\[\rho^v_i=\sum_{\left< i,j \right>\in {\cal P}(i)} \frac{1}{2 \pi a^2}({\bf z}\cdot{\bf S}_i\times {\bf S}_j), \]
where ${\cal P}(i)$ denotes all edges of plaquette $i$ and the ordering of edge indices corresponds to ${\bf r}_i-{\bf r}_j$ pointing along the counterclockwise walk around the plaquette (see Fig.~2 in the main text). In Fig.~\ref{fig:merons}, we plot a snapshot of magnetic texture containing $100\times 100$ magnetic moments. Color indicates the sign and magnitude of vorticity density.

\bibliography{lib}

\begin{thebibliography}{54}
\bibitem [{Not()}]{Note}%
  \BibitemOpen
  \href@noop {} {}\bibinfo {note} {For $J<0$, we obtain qualitatively similar
  results.}\BibitemShut {Stop}%
\end{thebibliography}

\end{document}